\documentclass[10pt, conference]{IEEEtran}
\IEEEoverridecommandlockouts

\usepackage{graphicx} %
\graphicspath{{./figures}}
\usepackage[OT1]{fontenc}
\usepackage{color}
\usepackage[dvipsnames,svgnames, table]{xcolor} %
\usepackage{siunitx}
\DeclareSIUnit{\dB}{dB}	                %
\DeclareSIUnit{\dBm}{dBm}	                %
\DeclareSIUnit{\dBi}{dBi}                   %
\DeclareSIUnit{\dBsm}{dBsm}                 %
\DeclareSIUnit{\ppm}{ppm}
\usepackage{eurosym}
\DeclareSIUnit{\EUR}{\text{\officialeuro}}
\DeclareUnicodeCharacter{20AC}{\officialeuro}
\usepackage{booktabs}
\usepackage[backend=biber, style=ieee, citestyle=numeric-comp,  minbibnames=99, maxbibnames=99,isbn=false, doi=false,citecounter=true]{biblatex}

\usepackage{circuitikz}

\usepackage{pgfplots}
\usepgfplotslibrary{groupplots,dateplot}
\usetikzlibrary{patterns,shapes.arrows}
\usetikzlibrary{spy}
\pgfplotsset{compat=newest}
\usetikzlibrary{external}
\usetikzlibrary{calc}
\usetikzlibrary{shadings}
\usetikzlibrary{shadows.blur}
\usetikzlibrary{decorations.pathreplacing}

\usepackage{soul}

\pgfmathdeclarefunction{viridis}{1}{%
  \pgfmathparse{%
    1000*(
    0.370489*pow(#1, 3)-
    1.11384*pow(#1, 2)+
    0.929777*(#1)
    )%
  }%
}

\usepackage{tabularx}
\usepackage{tablefootnote}

\AtBeginBibliography{\footnotesize}

\usepackage[font=small]{caption}
\usepackage{subcaption}
\usepackage{xspace}

\usepackage{placeins}
\usepackage[xindy, nonumberlist]{glossaries}
\makeglossaries%

\newacronym{2d}{2D}{two-dimensional}
\newacronym{3d}{3D}{three-dimensional}
\newacronym{3gpp}{3GPP}{3rd generation partnership project}
\newacronym{4g}{4G}{fourth-generation}
\newacronym{5g}{5G}{fifth-generation}
\newacronym{6dof}{6DoF}{six degrees of freedom}
\newacronym{6g}{6G}{sixth-generation}
\newacronym{abp}{ABP}{authentication by personalisation}
\newacronym{ac}{AC}{alternating current}
\newacronym{aclr}{ACLR}{adjacent channel leakage ratio}
\newacronym{adc}{ADC}{analog-to-digital converter}
\newacronym{adr}{ADR}{adaptive data rate}
\newacronym{aec}{AEC}{aluminum electrolytic capacitor}
\newacronym{aes}{AES}{Advanced Encryption Standard}
\newacronym{afe}{AFE}{analog front end}
\newacronym{ag}{AG}{array gain}
\newacronym{agc}{AGC}{automatic gain controller}
\newacronym{agps}{A-GPS}{Assisted Global Positioning System} %
\newacronym{ai}{AI}{artificial intelligence}
\newacronym{alk}{ALK}{Alkaline}
\newacronym{am}{AM}{amplitude modulation}
\newacronym{amam}{AM/AM}{amplitude modulation to amplitude modulation}
\newacronym{amc}{AMC}{automatic modulation classification}
\newacronym{amp}{AMP}{approximate message passing}
\newacronym{ampm}{AM/PM}{amplitude modulation to phase modulation}
\newacronym{aoa}{AOA}{angle-of-arrival}
\newacronym{aod}{AOD}{angle-of-departure}
\newacronym{ap}{AP}{access point}
\newacronym{api}{API}{application program interface}
\newacronym{apt}{APT}{acoustic power transfer}
\newacronym{apu}{APU}{access point unit}
\newacronym{ar}{AR}{augmented reality}
\newacronym{arima}{ARIMA}{Auto-Regressive Integrated Moving Average}
\newacronym{arp}{ARP}{Antenna Reference Point}
\newacronym{asic}{ASIC}{application specific integrated circuit}
\newacronym{ask}{ASK}{amplitude-shift keying}
\newacronym{at}{AT}{ATtention}
\newacronym{auv}{AUV}{autonomous underwater vehicle}
\newacronym{awg}{AWG}{arbitrary waveform generator}
\newacronym{awgn}{AWGN}{additive white Gaussian noise}
\newacronym{baw}{BAW}{bulk acoustic wave}
\newacronym{bb}{BB}{base-band}
\newacronym{bc}{BC}{backscatter communication}
\newacronym{bd}{BD}{backscatter device}
\newacronym{be}{BE}{Belgium}
\newacronym{ber}{BER}{bit error rate}
\newacronym{bf}{BF}{beamforming}
\newacronym{bga}{BGA}{ball grid array}
\newacronym{bldc}{BLDC}{brushless DC}
\newacronym{bod}{BOD}{Brown-Out Detection}
\newacronym{bom}{BOM}{bill of materials}
\newacronym{bpsk}{BPSK}{binary phase-shift keying}
\newacronym{brp}{BRP}{beam refinement process}
\newacronym{bs}{BS}{base station}
\newacronym{bu}{BU}{booster unit}
\newacronym{bw}{BW}{bandwidth}
\newacronym{ca}{CA}{carrier emitter}
\newacronym{cad}{CAD}{channel activity detection}
\newacronym{cars}{CARS}{calibration reference signal}
\newacronym{cbm}{CBM}{condition based maintenance}
\newacronym{cc}{CC}{constant current}
\newacronym{ccdf}{CCDF}{complementary cumulative distribution function}
\newacronym{ccnn}{CCNN}{circular convolutional neural network}
\newacronym{ccs}{CCS}{correlative channel sounder}
\newacronym{cdf}{CDF}{cumulative distribution function}
\newacronym{cdma}{CDMA}{code division-multiple access}
\newacronym{cdrx}{CDRX}{connected mode DRX}
\newacronym{ce}{CE}{coverage enhancement}
\newacronym{ced}{CED}{cumulative energy density}
\newacronym{cf}{CF}{cell-free}
\newacronym{cfmmimo}{CF-mMIMO}{cell-free massive MIMO}
\newacronym{cfo}{CFO}{carrier frequency offset}
\newacronym{cir}{CIR}{channel impulse response}
\newacronym{cla}{CLA}{closed-loop approach}
\newacronym{clk}{CLK}{clock}
\newacronym{cmos}{CMOS}{complementary metal oxide semiconductor}
\newacronym{cnn}{CNN}{convolutional neural network}
\newacronym{cordic}{CORDIC}{coordinate rotation digital computer}
\newacronym{cost}{COST}{commercial off-the-shelf}
\newacronym{cots}{COTS}{commercial off-the-shelf}
\newacronym{cp}{CP}{cyclic prefix}
\newacronym{cpt}{CPT}{capacitive power transfer}
\newacronym{cpu}{CPU}{central-processing unit}
\newacronym{cpw}{CPW}{coplanar waveguide}
\newacronym{cqi}{CQI}{channel quality indicator}
\newacronym{cr}{CR}{coding rate}
\newacronym{crc}{CRC}{cyclic redundancy check}
\newacronym{crlb}{CRLB}{Cram\'er-Rao lower bound}
\newacronym{crs}{CRS}{cell reference signal}
\newacronym{cs}{CS}{compressed sensing}
\newacronym{cse}{CSE}{channel state estimation}
\newacronym{csi}{CSI}{channel state information}
\newacronym{csp}{CSP}{contact service point}
\newacronym{css}{CSS}{chirp spread spectrum}
\newacronym{cu}{CU}{central unit}
\newacronym{cv}{CV}{constant voltage}
\newacronym{cw}{CW}{continuous wave}
\newacronym{d2d}{D2D}{device-to-device}
\newacronym{dab}{DAB}{Digital Audio Broadcasting}
\newacronym{dac}{DAC}{digital-to-analog converter}
\newacronym{daq}{DAQ}{data acquisition system}
\newacronym{das}{DAS}{distributed antenna systems}
\newacronym{dbpsk}{DBPSK}{Differential Binary Phase Shift Keying}
\newacronym{dc}{DC}{direct current}
\newacronym{dcc}{DCC}{dynamic cooperation clustering}
\newacronym{ddc}{DDC}{digital down conversion}
\newacronym{de}{DE}{drain efficiency}
\newacronym{dft}{DFT}{discrete Fourier transform}
\newacronym{dl}{DL}{downlink}
\newacronym{dlc}{DLC}{Distributed Laser Charging}
\newacronym{dli}{DLI}{direct link interference}
\newacronym{dlt}{DLT}{Distributed Ledger Technology}
\newacronym{dm}{DM}{diffuse multipath}
\newacronym{dma}{DMA}{Direct Memory Access}
\newacronym{dmac}{DMAC}{Direct Memory Access Controller}
\newacronym{dmc}{DMC}{diffuse multipath component}
\newacronym{dmimo}{D-MIMO}{distributed MIMO}
\newacronym{dnn}{DNN}{deep neural network}
\newacronym{doa}{DOA}{direction-of-arrival}
\newacronym{dp}{DP}{differencial privacy}
\newacronym{dpd}{DPD}{digital pre-distortion}
\newacronym{dpdk}{DPDK}{Data Plane Development Kit}
\newacronym{dram}{DRAM}{dynamic random-access memory}
\newacronym{drcs}{$\Delta$RCS}{differential-radar cross section}
\newacronym{drx}{DRX}{Discontinuous Reception Mode}
\newacronym{dsb}{DSB}{double-sideband}
\newacronym{dsl}{DSL}{digital subscriber line}
\newacronym{dsp}{DSP}{digital signal processing}
\newacronym{dss}{DSS}{Dataset Storage Standard}
\newacronym{duc}{DUC}{digital up-converter}
\newacronym{dvb}{DVB}{Digital Video Broadcasting}
\newacronym{e2e}{E2E}{end-to-end}
\newacronym{easa}{EASA}{European Union Aviation Safety Agency}
\newacronym{ebg}{EBG}{electromagnetic bandgap}
\newacronym{ec}{EC}{European Commission}
\newacronym{ecc}{ECC}{elliptic curve cryptography}
\newacronym{ecdf}{eCDF}{empirical cumulative distribution function}
\newacronym{ecdlp}{ECDLP}{elliptic curve discrete logarithm problem}
\newacronym{ecsp}{ECSP}{edge computing service point}
\newacronym{edlc}{EDLC}{electrostatic double-layer capacitors}
\newacronym{edrx}{eDRX}{Extended Discontinuous Reception Mode}
\newacronym{ee}{EE}{energy efficiency}
\newacronym{egprs}{EGPRS}{Enhanced Data Rates for GSM Evolution}
\newacronym{eh}{EH}{energy harvesting}
\newacronym{eirp}{EIRP}{equivalent isotropically radiated power}
\newacronym{em}{EM}{electromagnetic}
\newacronym{embb}{eMBB}{enhanced Mobile Broadband}
\newacronym{en}{EN}{energy neutral}
\newacronym{end}{END}{energy-neutral device}
\newacronym{enob}{ENOB}{effective number of bits}
\newacronym{eol}{EoL}{end of life}
\newacronym{ep}{EP}{energy profiler}
\newacronym{epd}{EPD}{electronic paper display}
\newacronym{epu}{EPU}{edge processing unit}
\newacronym{er}{ER}{Energy Receiver}
\newacronym{erp}{ERP}{effective radiation power}
\newacronym[plural=ESCs,firstplural=electronic speed controllers (ESCs)]{esc}{ESC}{electronic speed control}
\newacronym{esd}{ESD}{electrostatic discharge}
\newacronym{esl}{ESL}{electronic shelf label}
\newacronym{esr}{ESR}{equivalent series resistance}
\newacronym{et}{ET}{Energy Transmitter}
\newacronym{etsi}{ETSI}{European Telecommunications Standards Institute}
\newacronym{evd}{EVD}{eigenvalue decomposition}
\newacronym{evm}{EVM}{Error Vector Magnitude}
\newacronym{ewlb}{eWLB}{Embedded Wafer Level Ball Grid Array}
\newacronym{fa}{FA}{federation anchor}
\newacronym{fair}{FAIR}{Findability, Accessibility, Interoperability, and Reuse of digital assets}
\newacronym{fcf}{FCF}{frequency correlation function}
\newacronym{fd}{FD}{front-haul distance}
\newacronym{fdd}{FDD}{frequency-division duplexing}
\newacronym{fde}{FDE}{frequency domain equalizer}
\newacronym{fdm}{FDM}{frequency-division multiplexing}
\newacronym{fdma}{FDMA}{frequency division multiple access}
\newacronym{fem}{FEM}{finite element analysis}
\newacronym{fembb}{feMBB}{further enhanced mobile broadband}
\newacronym{fft}{FFT}{fast Fourier transform}
\newacronym{fh}{FH}{fronthaul}
\newacronym{fhss}{FHSS}{frequency hopping spread spectrum}
\newacronym{fifo}{FIFO}{First In, First Out}
\newacronym{fim}{FIM}{Fisher information matrix}
\newacronym{fits}{FITS}{Flexible Image Transport System}
\newacronym{fl}{FL}{federated learning}
\newacronym{fom}{FoM}{figure of merit}
\newacronym{fov}{FOV}{field of view}
\newacronym{fpga}{FPGA}{field-programmable gate array}
\newacronym{fram}{FRAM}{Ferroelectric Random Access Memory}
\newacronym{fsk}{FSK}{frequency shift keying}
\newacronym{fspl}{FSPL}{free space path loss}
\newacronym{fss}{FSS}{frequency selective surface}
\newacronym{gb}{GB}{grant-based}
\newacronym{gdpr}{GDPR}{general data protection regulation}
\newacronym{gf}{GF}{grant-free}
\newacronym{gmsk}{GMSK}{Gaussian minimum-shift keying}
\newacronym{gnb}{gNB}{Next Generation Node B}
\newacronym{gni}{GNI}{gross national income}
\newacronym{gnn}{GNN}{graph neural network}
\newacronym{gnss}{GNSS}{global navigation satellite system}
\newacronym{gpclk}{GPCLK}{general purpose clock}
\newacronym{gpio}{GPIO}{General-Purpose Input/Output}
\newacronym{gpl}{GPL}{GNU General Public License}
\newacronym{gprs}{GPRS}{General Packet Radio Services}
\newacronym{gps}{GPS}{Global Positioning System}
\newacronym{gpu}{GPU}{graphical processing unit}
\newacronym{grc}{GRC}{GNU Radio Companion}
\newacronym{gscm}{GSCM}{geometry‐based stochastic model}
\newacronym{gsm}{GSM}{Global System for Mobile Communications}  %
\newacronym{gwp}{GWP}{Global Warming Potential}
\newacronym{harq}{HARQ}{hybrid automatic repeat request}
\newacronym{hat}{HAT}{hardware attached on top}
\newacronym{hcs}{HCS}{human-centric services}
\newacronym{hdf5}{HDF5}{Hierarchical Data Format version 5}
\newacronym{hfss}{HFSS}{High Frequency Simulator Software}
\newacronym{hmd}{HMD}{head-mounted display}
\newacronym{hpbm}{HPBM}{half power beam width}
\newacronym{HyMPRo}{HyMPRo}{Hybrid Multi-Path Routing algorithm}
\newacronym{i.i.d.}{i.i.d.}{independent and identically distributed}
\newacronym{i2c}{I2C}{Inter-Integrated Circuit}
\newacronym{iaq}{IAQ}{Indoor Air Quality}
\newacronym{ib}{IB}{in-band}
\newacronym{ibbc}{IBBC}{inter-band beam configuration}
\newacronym{ibo}{IBO}{input back-off}
\newacronym{ic}{IC}{integrated circuit}
\newacronym{iccs}{ICCS}{Ilmsens correlative channel sounder}
\newacronym{ici}{ICI}{intercarrier interference}
\newacronym{icnirp}{ICNIRP}{International Commission on Non-Ionizing Radiation Protection}
\newacronym{id}{ID}{information decoding}
\newacronym{idft}{IDFT}{inverse discrete Fourier transform}
\newacronym{if}{IF}{intermediate-frequency}
\newacronym{iid}{i.i.d.}{independently and identically distributed}
\newacronym{iis}{IIS}{integrated information system}
\newacronym{im}{IM}{intermodulation}
\newacronym{imd}{IMD}{intermodulation distortion}
\newacronym{imu}{IMU}{inertial measurement unit}
\newacronym{inh}{InH}{indoor hotspot office}
\newacronym{io}{IO}{input/output}
\newacronym{ioe}{IoE}{Internet of Everything}
\newacronym{iot}{IoT}{Internet of Things}
\newacronym{ipt}{IPT}{inductive power transfer}
\newacronym{ipy}{IPY}{Interventions per Year}
\newacronym{iq}{IQ}{in-phase and quadrature}
\newacronym{iqi}{IQI}{IQ imbalance}
\newacronym{ir}{IR}{infrared}
\newacronym{isi}{ISI}{intersymbol interference}
\newacronym{ism}{ISM}{industrial, scientific and medical}
\newacronym{isp}{ISP}{internet service provider}
\newacronym{itu}{ITU}{International Telecommunication Union}
\newacronym{jesd}{JESD}{Joint Electron Devices Engineering Council}
\newacronym{jfet}{JFET}{junction field effect transistor}
\newacronym{kpi}{KPI}{key performance indicator}
\newacronym{kvi}{KVI}{key value indicator}
\newacronym{larva}{LARVA}{LARge Virtual Array}
\newacronym{lca}{LCA}{life cycle assessment}
\newacronym{lco}{LCO}{lithium cobalt oxide}
\newacronym{ldo}{LDO}{Low-dropout voltage regulator}
\newacronym{ldpc}{LDPC}{low-density parity-check}
\newacronym{led}{LED}{Light Emitting Diode}
\newacronym{less}{LESS}{Low Energy Scheduler Solution}
\newacronym{lfp}{LFP}{lithium iron phosphate}
\newacronym{lib}{LIB}{Lithium-Ion Battery}
\newacronym{lic}{LIC}{lithium-ion capacitor}
\newacronym{lid}{LID}{Lithium Iron Disulfide}
\newacronym{lidar}{LiDAR}{light detection and ranging}
\newacronym{liion}{Li-ion}{lithium-ion}
\newacronym{lipo}{LiPo}{lithium polymer}
\newacronym{lis}{LIS}{large intelligent surface}
\newacronym{llh}{LLH}{log-likelihood}
\newacronym{lmd}{LMD}{Lithium Manganese Dioxide}
\newacronym{lmmse}{LMMSE}{least minimum mean square error}
\newacronym{lmo}{LMO}{lithium ion manganese oxide}
\newacronym{lna}{LNA}{low-noise amplifier}
\newacronym{lo}{LO}{local oscillator}
\newacronym{lora}{LoRa}{long range}
\newacronym{lorawan}{LoRaWAN}{long-range wide-area network}
\newacronym{los}{LoS}{line-of-sight}
\newacronym{lp}{LP}{linear programming}
\newacronym{lpf}{LPF}{low-pass filter}
\newacronym{lpt}{LPT}{laser power transfer}
\newacronym{lpwa}{LPWA}{Low Power Wide Area}
\newacronym{lpwan}{LPWAN}{low-power wide-area network}
\newacronym{lpwans}{LPWANs}{Low-Power Wide-Area Networks}
\newacronym{lqi}{LQI}{link quality indicator}
\newacronym{lrelu}{LReLU}{leaky rectified linear unit}
\newacronym{lrt}{LRT}{likelihood-ratio test}
\newacronym{ls}{LS}{least squares}
\newacronym{lsa}{LSA}{large synthetic array}
\newacronym{lsf}{LSF}{large-scale fading}
\newacronym{lsfc}{LSFC}{large-scale fading component}
\newacronym{lstm}{LSTM}{Long Short-Term Memory}
\newacronym{ltc}{LTC}{lithium thionyl chloride}
\newacronym{lte}{LTE}{Long Term Evolution}
\newacronym{lti}{LTI}{linear time-invariant}
\newacronym{lto}{LTO}{lithium titanate}
\newacronym{lusta}{LUSTA 5G }{Logistique mUltimodale Sécuritaire Téléopérée \& Autonome 5G}
\newacronym{m2m}{M2M}{machine to machine}
\newacronym{mac}{MAC}{Medium Access Control}
\newacronym{mate}{MATE}{millimeter-wave MIMO testbed}
\newacronym{mc}{MC}{Monte Carlo}
\newacronym{mcl}{MCL}{Maximum Coupling Loss}
\newacronym{mcs}{MCS}{modulation and coding scheme}
\newacronym{mcu}{MCU}{microcontroller unit}
\newacronym{mec}{MEC}{multi-access edge computing}
\newacronym{mems}{MEMS}{micro-electromechanical systems}
\newacronym{mf}{MF}{matched filter}
\newacronym{mimo}{MIMO}{multiple-input multiple-output}
\newacronym{miso}{MISO}{multiple-input single-output}
\newacronym{ml}{ML}{machine learning}
\newacronym{mlp}{MLP}{multilayer perceptron}
\newacronym{mmic}{MMIC}{monolithic microwave integrated circuit}
\newacronym{mmimo}{mMIMO}{massive MIMO}
\newacronym{mmse}{MMSE}{minimum mean square error}
\newacronym{mmtc}{mMTC}{massive machine-typed communication}
\newacronym{mmwave}{mmWave}{millimeter wave}
\newacronym{mn}{MN}{matching network}
\newacronym{mosfet}{MOSFET}{metal-oxide semiconductor field effect transistor}
\newacronym{mpc}{MPC}{multipath component}
\newacronym{mppt}{MPPT}{maximum power point tracking}
\newacronym{mr}{MR}{maximum ratio}
\newacronym{mrc}{MRC}{maximum ratio combining}
\newacronym{mrc_em}{MRC}{maximum ratio combining}
\newacronym{mrc_EM}{MRC}{Magnetic Resonance Coupling}
\newacronym{mrt}{MRT}{maximum ratio transmission}
\newacronym{mse}{MSE}{mean square error}
\newacronym{msk}{MSK}{Minimum-Shift Keying}
\newacronym{mtc}{MTC}{Machine-Type Communication}
\newacronym{multi-rat}{Multi-RAT}{multiple radio access technology}
\newacronym{multirat}{Multi-RAT}{Multiple Radio Access Technology}
\newacronym{music}{MUSIC}{MUltiple SIgnal Classification}
\newacronym{navauwall}{NAVAUWALL}{AUtomated NAVigation in WALLonia}
\newacronym{nb}{NB}{narrowband}
\newacronym{nbiot}{NB-IoT}{narrowband IoT}
\newacronym{nca}{NCA}{nickel cobalt aluminum}
\newacronym{netcdf}{NetCDF}{Network Common Data Form}
\newacronym{nf}{NF}{noise figure}
\newacronym{nfc}{NFC}{near-field communication}
\newacronym{nfv}{NFV}{network function virtualization}
\newacronym{ngmn}{NGMN}{Next Generation Mobile Networks }
\newacronym{ni}{NI}{National Instruments}
\newacronym{nicd}{NiCd}{nikkel cadmium}
\newacronym{nimh}{NiMH}{nikkel metal hydride}
\newacronym{nlos}{NLoS}{non-line-of-sight}
\newacronym{nmc}{NMC}{nickel manganese cobalt}
\newacronym{nmos}{nMOS}{n-channel metal-oxide semiconductor}
\newacronym{nn}{NN}{neural network}
\newacronym{nnls}{NNLS}{non-negative least squares}
\newacronym{noma}{NOMA}{non-orthogonal multiple access}
\newacronym{np}{NP}{Neyman-Pearson}
\newacronym{npbch}{NPBCH}{Narrowband Physical Broadcast Channel}
\newacronym{npss}{NPSS}{Narrow Band Primay Synchronization Signal}
\newacronym{nr}{NR}{New Radio}
\newacronym{nrs}{NRS}{Narrow Band Reference Signal}
\newacronym{nsss}{NSSS}{Narrowband Secondary Synchronization Signal}
\newacronym{ntp}{NTP}{network time protocol}
\newacronym{oai}{OAI}{OpenAirInterface} %
\newacronym{obw}{OBW}{occupied bandwidth}
\newacronym{ofdm}{OFDM}{orthogonal frequency-division multiplexing}
\newacronym{ofdma}{OFDMA}{orthogonal frequency-division multiple access}
\newacronym{ofdmim}{OFDM-IM}{OFDM with index modulation}
\newacronym{olos}{OLoS}{obstructed-line-of-sight}
\newacronym{oma}{OMA}{orthogonal multiple access}
\newacronym{oob}{OOB}{out-of-band}
\newacronym{ook}{OOK}{on-off keying}
\newacronym{oran}{O-RAN}{open radio-access network}
\newacronym{os}{OS}{operating system}
\newacronym{ota}{OTA}{over-the-air}
\newacronym{otaa}{OTAA}{over-the-air authentication}
\newacronym{p1}{P1}{Phase 1}
\newacronym{p2}{P2}{Phase 2}
\newacronym{p2p}{P2P}{point-to-point}
\newacronym{pa}{PA}{power amplifier}
\newacronym{pae}{PAE}{power-added efficiency}
\newacronym{pana}{PanA}{Panel A}
\newacronym{panb}{PanB}{Panel B}
\newacronym{papr}{PAPR}{peak-to-average power ratio}
\newacronym{pc}{PC}{pilot count}
\newacronym{pcb}{PCB}{printed circuit board}
\newacronym{pcg}{PCG}{power consumption gain}
\newacronym{pcie}{PCIe}{Peripheral Component Interconnect Express}
\newacronym{pcsi}{PCSI}{perfect channel state information}
\newacronym{pd}{PD}{powered device}
\newacronym{pdcch}{PDCCH}{physical downlink control channel}
\newacronym{pdf}{PDF}{probability density function}
\newacronym{pdp}{PDP}{power delay profile}
\newacronym{pdsch}{PDSCH}{physical downlink shared channel}
\newacronym{pe}{PE}{processing element}
\newacronym{peb}{PEB}{positioning error bound}
\newacronym{per}{PER}{packet error rate}
\newacronym{pet}{PET}{privacy enhancing technology}
\newacronym{pg}{PG}{path gain}
\newacronym{pgd}{PGD}{proximal gradient descent}
\newacronym{phy}{PHY}{physical}
\newacronym{pki}{PKI}{public key infrastucture}
\newacronym{pl}{PL}{path loss}
\newacronym{pla}{PLA}{physically large array}
\newacronym{pll}{PLL}{phase-locked loop}
\newacronym{pmf}{PMF}{polymer microwave fiber}
\newacronym{pmu}{PMU}{power management unit}
\newacronym{pn}{PN}{pseudo-noise}
\newacronym{poc}{PoC}{proof of concept}
\newacronym{poe}{PoE}{power-over-Ethernet}
\newacronym{pps}{1PPS}{pulse per second}
\newacronym{pr}{PR}{phase reversal}
\newacronym{prb}{PRB}{Physical Resource Block}
\newacronym{prbs}{PRBs}{Physical Resource Blocks}
\newacronym{prs}{PRS}{Peripheral Reflex System}
\newacronym{ps}{PS}{Processing System}
\newacronym{psd}{PSD}{power spectral density}
\newacronym{pse}{PSE}{power sourcing equipment}
\newacronym{psk}{PSK}{phase shift keying}
\newacronym{psm}{PSM}{power saving mode}
\newacronym{pss}{PSS}{primary synchronisation signal}
\newacronym{ptp}{PTP}{precision-time protocol}
\newacronym{ptrs}{PTRS}{Phase-Tracking Reference Signals}
\newacronym{ptw}{PTW}{paging time window}
\newacronym{pv}{PV}{photovoltaic}
\newacronym{pw}{PW}{planar wavefront}
\newacronym{pwm}{PWM}{pulse width modulation}
\newacronym{qam}{QAM}{quadrature amplitude modulation}
\newacronym{qos}{QoS}{quality-of-service}
\newacronym{qpsk}{QPSK}{quadrature phase-shift keying}
\newacronym{qrrls}{QR-RLS}{QR decomposition based recursive least squares}
\newacronym{quadriga}{QuaDRiGa}{QUAsi Deterministic RadIo channel GenerAtor}
\newacronym{ra}{RA}{Random Access}
\newacronym{ram}{RAM}{random-access memory}
\newacronym{ran}{RAN}{radio access network}
\newacronym{rar}{RAR}{Random Access Response}
\newacronym{rat}{RAT}{radio access technology}
\newacronym{rb}{Rb}{Rubidium}
\newacronym{rbs}{RBS}{radio base station}
\newacronym{rbw}{RBW}{resolution bandwidth}
\newacronym{rcs}{RCS}{radar cross section}
\newacronym{rdl}{RDL}{redistribution layer}
\newacronym{re}{RE}{radio element}
\newacronym{relu}{ReLU}{rectified linear unit}
\newacronym{rf}{RF}{radio frequency}
\newacronym{rfeh}{RFEH}{radio frequency energy harvesting}
\newacronym{rfic}{RFIC}{radio-frequency integrated circuit}
\newacronym{rfid}{RFID}{radio frequency identification}
\newacronym{rfpt}{RFPT}{radio frequency power transfer}
\newacronym{rfsoc}{RFSoC}{Radio Frequency System-on-Chip}
\newacronym{rir}{RIR}{room impulse response}
\newacronym{ris}{RIS}{reflective intelligent surface}%
\newacronym{rllmtc}{RLLMTC}{reliable low latency machine type communication}
\newacronym{rls}{RLS}{recursive least squares}
\newacronym{rms}{RMS}{root-mean-square}
\newacronym{rmse}{RMSE}{root-mean-square error}
\newacronym{rmt}{RMT}{random matrix theory}
\newacronym{rof}{RoF}{radio-over-fiber}
\newacronym{ros}{ROS}{robot operating system}
\newacronym{rpi}{RPi}{Raspberry Pi}
\newacronym{rrc}{RRC}{Radio Resource Connection}
\newacronym{rreq}{RREQ}{route request packet}
\newacronym{rsrp}{RSRP}{Reference Signals Received Power}
\newacronym{rsrq}{RSRQ}{Reference Signal Received Quality}
\newacronym{rss}{RSS}{received signal strength}
\newacronym{rssi}{RSSI}{received signal strength indicator}
\newacronym{rtc}{RTC}{real time clock}
\newacronym{rtf}{RTF}{reader talks first}
\newacronym{rtk}{RTK}{real time kinematics}
\newacronym{rts}{RTS}{ray tracing simulator}
\newacronym{ru}{RU}{Radio Unit}
\newacronym{rv}{RV}{random variable}
\newacronym{rw}{RW}{RadioWeaves}
\newacronym{rx}{RX}{receiver}
\newacronym{rzf}{RZF}{regularized zero forcing}
\newacronym{s-parameter}{S-parameter}{scattering parameter}
\newacronym{sa}{SA}{synchronization anchor}
\newacronym{sa5}{SA}{Stand Alone}
\newacronym{sar}{SAR}{specific absorption rate}
\newacronym{sbl}{SBL}{sparse Bayesian learning}
\newacronym{scfdma}{SCFDMA}{single-carrier frequency division multiple access}
\newacronym{sdg}{SDG}{Sustainable Development Goal}
\newacronym{sdm}{SDM}{sigma-delta modulator}
\newacronym{sdma}{SDMA}{spatial-division multiple access}
\newacronym{sdn}{SDN}{software-defined network}
\newacronym{sdof}{SDoF}{sigma-delta over fiber}
\newacronym{sdr}{SDR}{software-defined radio}
\newacronym{se}{SE}{spectral efficiency}
\newacronym{sei}{SEI}{specific emitter identification}
\newacronym{ser}{SER}{symbol-error rate}
\newacronym{sf}{SF}{spreading factor}
\newacronym{sfn}{SFN}{single frequency network}
\newacronym{sfp}{SFP}{small form-factor pluggable}
\newacronym{sha}{SHA}{Secure Hash Algorithm}
\newacronym{SigMF}{SigMF}{Signal Metadata Format}
\newacronym{simo}{SIMO}{single-input multiple-output}
\newacronym{sinr}{SINR}{signal-to-interference-plus-noise ratio}
\newacronym{siso}{SISO}{single-input single-output}
\newacronym{slam}{SLAM}{simultaneous localization and mapping}
\newacronym{slc}{SLC}{spatial leakage suppression}
\newacronym{slerp}{SLERP}{spherical linear interpolation}
\newacronym{sma}{SMA}{SubMiniature version A}
\newacronym{smc}{SMC}{specular multipath component}
\newacronym{smps}{SMPS}{switched mode power supply}
\newacronym{sndr}{SNDR}{signal-to-noise-and-distortion ratio}
\newacronym{snidr}{SNIDR}{signal-to-noise-and-interference-and-distortion ratio}
\newacronym{snir}{SNIR}{signal-to-interference-plus-noise ratio}
\newacronym{snr}{SNR}{signal-to-noise ratio}
\newacronym{soc}{SoC}{state of charge}
\newacronym{SoC}{SOC}{System on Chip}
\newacronym{sp}{SP}{service point}
\newacronym{spdt}{SPDT}{single pole double throw}
\newacronym{spi}{SPI}{Serial Peripheral Interface}
\newacronym{spst}{SPST}{single pole single throw}
\newacronym{sram}{SRAM}{static random-access memory}
\newacronym{srd}{SRD}{short-range device}
\newacronym{srls}{SRLS}{standard recursive least squares}
\newacronym{srs}{SRS}{Sounding Reference Signal}
\newacronym{ssb}{SSB}{synchronisation signal block}
\newacronym{ssd}{SSD}{solid state drive}
\newacronym{ssq}{SSQ}{simulator sickness questionnaire}
\newacronym{steam}{STEAM}{science, technology, engineering, the arts, and mathematics}
\newacronym{svd}{SVD}{singular value decomposition}
\newacronym{sw}{SW}{spherical wavefront}
\newacronym{swipt}{SWIPT}{simultaneous wireless information and power transfer}
\newacronym{synce}{SyncE}{Synchronous Ethernet}
\newacronym{ta}{TA}{timing advance}
\newacronym{tau}{TAU}{tracking area update}
\newacronym{tcer}{TCER}{transported to consumed energy ratio}
\newacronym{tcp}{TCP}{Transmission Control Protocol}
\newacronym{tcxo}{TCXO}{temperature compensated crystal oscillator}
\newacronym{tdd}{TDD}{time division duplexing}
\newacronym{tdma}{TDMA}{time division-multiple access}
\newacronym{tdoa}{TDOA}{time-difference-of-arrival}
\newacronym{thz}{THz}{Terahertz}
\newacronym{to}{TO}{timing offset}
\newacronym{toa}{TOA}{time-of-arrival}
\newacronym{tof}{ToF}{time-of-flight}
\newacronym{tosm}{TOSM}{through-open-short-match}
\newacronym{tpms}{TPMS}{Tire-Pressure Monitoring System}
\newacronym{trl}{TRL}{technology readyness level}
\newacronym{trp}{TRP}{Transmission Reception Point}
\newacronym{tsn}{TSN}{time-sensitive networking}
\newacronym{ttf}{TTF}{tag talks first}
\newacronym{ttff}{TTFF}{Time To First Fix}
\newacronym{ttm}{TTM}{time to market}
\newacronym{ttn}{TTN}{The Things Network}
\newacronym{tx}{TX}{transmitter}
\newacronym{uart}{UART}{Universal Asynchronous Receiver/Transmitter}
\newacronym{uav}{UAV}{unmanned aerial vehicle}
\newacronym{uc}{UC}{use case}
\newacronym{ucie}{UCIe}{Universal Chiplet Interconnect Express}
\newacronym{udp}{UDP}{User Datagram Protocol}
\newacronym{ue}{UE}{user equipment}
\newacronym{ugv}{UGV}{unmanned ground vehicle}
\newacronym{uhd}{UHD}{USRP hardware driver}
\newacronym{uhf}{UHF}{ultra-high frequency}
\newacronym{ul}{UL}{uplink}
\newacronym{ula}{ULA}{uniform linear array}
\newacronym{uMIMO}{$\mu$-MIMO}{ultra-massive Multiple-Input Multiple-Output}
\newacronym{ummtc}{umMTC}{ultra massive machine type communication}
\newacronym{un}{UN}{United Nations}
\newacronym{upa}{UPA}{uniform planar array}
\newacronym{ura}{URA}{uniform rectangular array}
\newacronym{urllc}{URLLC}{ultra-reliable low-latency communications}
\newacronym{usrp}{USRP}{universal software radio peripheral}
\newacronym{uv}{UV}{unmanned vehicle}
\newacronym{uwb}{UWB}{ultrawideband}
\newacronym{v2v}{V2V}{vehicle-to-vehicle}
\newacronym{vco}{VCO}{voltage-controlled oscillator}
\newacronym{vep}{VEP}{virtual edge platform}
\newacronym{vlc}{VLC}{visible light communication}
\newacronym{vlp}{VLP}{visible light positioning}
\newacronym{vna}{VNA}{vector network analyzer}
\newacronym{voc}{VOC}{Voltatile Organic Compound}
\newacronym{votable}{VOTable}{Virtual Observatory Table}
\newacronym{vr}{VR}{virtual reality}
\newacronym{vuca}{VUCA}{volatile, uncertain, complex and ambiguous}
\newacronym{wb}{WB}{wideband}
\newacronym{wban}{WBAN}{wireless body area network}
\newacronym{wd}{WD}{Wireless distance}
\newacronym{wimax}{WiMAX}{Worldwide Interoperability for Microwave Access}
\newacronym{wlan}{WLAN}{wireless LAN}
\newacronym{wpt}{WPT}{wireless power transfer}
\newacronym{wr}{WR}{White Rabbit}
\newacronym{wrsn}{WRSN}{wireless rechargeable sensor network}
\newacronym{wsn}{WSN}{wireless sensor network}
\newacronym{xets}{XETS}{cross exponentially tapered slot}
\newacronym{xlmimo}{XL-MIMO}{extremely large-scale MIMO}
\newacronym{xr}{XR}{extended reality}
\newacronym{z3ro}{Z3RO}{zero third-order distortion}
\newacronym{zf}{ZF}{zero-forcing}
\newacronym{zmcscg}{ZMCSCG}{zero mean circularly symmetric complex Gaussian}
\newacronym{zmq}{ZMQ}{ZeroMQ}

\usepackage{comment}

\newcolumntype{R}{>{\raggedleft\arraybackslash}X}

\newcolumntype{H}{>{\setbox0=\hbox\bgroup}c<{\egroup}@{}}

\usepackage{graphicx}
\usepackage{textcomp}
\usepackage{xcolor}
\def\BibTeX{{\rm B\kern-.05em{\sc i\kern-.025em b}\kern-.08em
    T\kern-.1667em\lower.7ex\hbox{E}\kern-.125emX}}

\definecolor{darkgray176}{RGB}{176,176,176}

\usepackage[capitalize]{cleveref}

\usepackage{fontawesome}

\pgfplotsset{
    every axis/.append style={
        legend style={
            font=\small,       %
            nodes={scale=0.8, transform shape}
        },
         xlabel style={
            font=\small        %
        },
        ylabel style={
            font=\small        %
        },
    },
    every tick label/.append style={font=\small}
}

\tikzset{
    every node/.append style={font=\footnotesize},
    every label/.append style={font=\footnotesize}
}
\pgfplotsset{
    every axis legend/.append style={
        fill opacity=0.8,
        draw=none,
    },
    every axis/.append style={
        x grid style={darkgray!60},
        xmajorgrids,
        y grid style={darkgray!60},
        ymajorgrids,
    }
}

\usepackage{balance}

\input{auto_names}

\usepackage[normalem]{ulem}

\def\BibTeX{{\rm B\kern-.05em{\sc i\kern-.025em b}\kern-.08em
    T\kern-.1667em\lower.7ex\hbox{E}\kern-.125emX}}

\defineauthors[false]{jarne,gilles,liesbet,lieven,benjamin,thomas,todo,geof, reviewcomment, update}

\newcommand{\req}[1]{\textcolor{RedOrange}{R#1}}

\usepackage{tabularray}

\begin{document}

\title{%
Designing RF-Powered Battery-Less Electronic Shelf Labels With COTS Components \thanks{%
The work is partially supported by the REINDEER and AMBIENT-6G project under grant agreements No.~101013425 and No~101192113, respectively, as well as by the e-construct project under the 'Vlaams Agentschap Innoveren en Ondernemen (VLAIO)' grant agreement HBC.2021.0911.}
}

\author{
    \IEEEauthorblockN{Jarne Van Mulders and Gilles Callebaut}
    \IEEEauthorblockA{\textit{KU Leuven,} Belgium}
}

\maketitle

\begin{abstract}

This paper presents a preliminary study exploring the feasibility of designing batteryless \glspl{esl} powered by radio frequency wireless power transfer using \acrlong{cots} components. The proposed \gls{esl} design is validated through a dedicated testbed and involves a detailed analysis of design choices, including energy consumption, energy conversion, and storage solutions. A leaded aluminium electrolytic capacitor is selected as the primary energy storage element, balancing cost and performance while maintaining compactness. Experimental evaluations demonstrate that an \gls{esl} can update its display within 4 to 120 minutes, depending on input power and RF frequency, with harvester efficiencies reaching up to \SI{30}{\percent}. Challenges such as low harvester efficiency, extended update times, and hardware constraints are identified, highlighting opportunities for future optimizations. This work provides valuable insights into system design considerations for RF-powered \glspl{esl} and establishes a foundation for further research in energy-neutral \acrlong{iot} applications.

\end{abstract}


\glsresetall
\section{Introduction}
Recent studies and dedicated projects have focused on \glspl{end}~\cite{reindeer, hexax, ambient6g}, making it a key topic in ongoing 6G research. One specific use case is \glspl{esl}, which have low energy and reliability requirements and are produced in large quantities. Currently, different market players implement their own power and communication solutions for these \glspl{esl}, which are proprietary. This paper investigates a potential implementation while identifying improvements and trade-offs through experiments focused on energy consumption, conversion, and storage. It offers a comprehensive overview of key considerations and design guidelines for developing such devices.

\textbf{Device Classes. } This work focuses on Class~2 devices, as defined in \cite[Sec.~3.1]{ReindeerD11}, to address the specific challenges of \glspl{esl}. Reindeer~\cite{ReindeerD11} categorized devices into five classes based on energy supply and communication capabilities. Class~1 devices, relying solely on \gls{rf} energy through \gls{wpt} and backscatter communication, and are unsuitable due to the buffering required for energy-intensive screen updates. Conversely, Class~3 devices, commonly used in commercial \glspl{esl}, rely on batteries or wireless charging, which this study aims to avoid. Class~2 devices, being energy-neutral and battery-free, are ideal for updating \glspl{epd} in low-power scenarios.

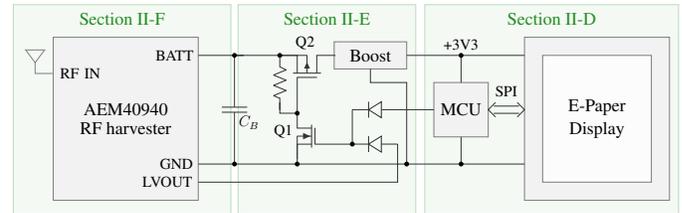
\begin{figure}[htbp]
    \centering
    \resizebox{\linewidth}{!}{\begin{tikzpicture}[american voltages, every text node part/.style={align=center}, /tikz/circuitikz/tripoles/nigfetd/height=.8, /tikz/circuitikz/tripoles/nigfetd/width=.5]

\ctikzset{bipoles/resistor/height=0.25}
\ctikzset{bipoles/resistor/width=0.6}
\ctikzset{bipoles/capacitor/height=0.5}
\ctikzset{bipoles/capacitor/width=0.1}
\ctikzset{bipoles/americaninductor/height=0.35}
\ctikzset{bipoles/americaninductor/width=0.8}

\definecolor{colorSections}{rgb}{0.13, 0.55, 0.13}

\tikzset{block/.style = {rectangle, draw=black!50, fill=black!5, thick, minimum width=4cm, minimum height = 4.5cm}}
\tikzset{boost/.style = {rectangle, draw=black!50, fill=black!5, thick, minimum width=2cm, minimum height = 0.75cm}}
\tikzset{mcu/.style = {rectangle, draw=black!50, fill=black!5, thick, minimum width=1.5cm, minimum height = 1.5cm}}
\tikzset{edp-screen/.style = {rectangle, draw=black!50, fill=black!1, thick, minimum width=3cm, minimum height = 3.5cm}}

\node[rectangle, draw=colorSections!30, fill=colorSections!5, thick, minimum width=7cm, minimum height = 5.8cm, text centered, text depth = 5 cm] at (12.5,2.0) () {\Large \color{colorSections} \cref{sec:mcu-epd}};
s
\node[rectangle, draw=colorSections!30, fill=colorSections!5, thick, minimum width=4.9cm, minimum height = 5.8cm, text centered, text depth = 5 cm] at (6.3,2.0) () {\Large \color{colorSections} \cref{sec:cap-storage}};

\node[rectangle, draw=colorSections!30, fill=colorSections!5, thick, minimum width=6cm, minimum height = 5.8cm, text centered, text depth = 5 cm] at (0.65,2.0) () {\Large \color{colorSections} \cref{sec:harvester}};

\node [block] at (0.75,1.75) () {\Large AEM40940\\[5pt] \Large RF harvester \par};
\node [boost] at (7.5,3.5) () { \Large Boost};
\node [mcu] at (10,2) () { \Large MCU};
\node [block] at (13.75,1.75) () {};
\node [edp-screen] at (13.75,1.75) () {\Large E-Paper\\[5pt] \Large Display \par};

\node [] at (10,3.8) () {\large +3V3};
\node [] at (11.25,2.5) () {\large SPI};

\draw[color=black!80]
  (4,3.5) -- (3.75,3.5) to[C] (3.75,0.5) %
  (2.75,3.5) -- (5,3.5)
  (6.5,3.5) to [Tpigfetd,n=pmos1] (5,3.5)
  (5,3.5) to[R] (5,2)
  ($(pmos1.G) - (0,2)$) to [Tnigfetd,n=nmos1,mirror,/tikz/circuitikz/bipoles/length=45pt] ($(pmos1.G) - (0,0.5)$)
  (pmos1.G) -- +(0,-0.5)
  ($(pmos1.G) - (0,2)$) |- (2.75,0.5)
  (5,2.2) |- (nmos1.D) 
  (nmos1.G) -| (7,2) 
  (9.25,2) -- (8.25,2) to[D,/tikz/circuitikz/bipoles/length=25pt] (7,2)
  (8.25,1.05) to[D,/tikz/circuitikz/bipoles/length=25pt] (7,1.05) 
  (8.25,1.05) |- (2.75,0)
  (8.5,3.5) -- (11.75,3.5)
  (10,2.75) -- (10,3.5)
  (10.9,2.1) -- (11.6,2.1)
  (10.9,1.9) -- (11.6,1.9)
  (10.95,2.2) -- (10.75,2) --  (10.95,1.8)
  (11.55,2.2) -- (11.75,2) --  (11.55,1.8)
  ($(pmos1.G) - (0,2)$) |- (10,0.5) -- (10,1.25)
  (10,0.5) -- (11.75,0.5)
  (8.5,0.5) -- (8.5,2.75) -- (7.5,2.75) -- (7.5,3.125) 
;

\draw[draw=black!50]
(-1.25,3) -- (-1.75,3)-- +(0,0.3) -- +(0.2625,0.65) -- +(-0.2625,0.65) -- +(0,0.3)
;

\node [] at (2.1,3.5) () {\large BATT};
\node [] at (2.15,0.5) () {\large GND};
\node [] at (1.9,0) () {\large LVOUT};
\node [] at (-0.5,3) () {\large RF IN};
\node [] at (5.1,1.4) () {\large Q1};
\node [] at (5.7,3.85) () {\large Q2};
\node [] at (4.15,1.65) () {\large $C_{B}$};

\filldraw[black] (nmos1.D) circle (1.5pt) node[anchor=west]{};
\filldraw[black] (3.75,0.5) circle (1.5pt) node[anchor=west]{};
\filldraw[black] (3.75,3.5) circle (1.5pt) node[anchor=west]{};
\filldraw[black] (10,3.5) circle (1.5pt) node[anchor=west]{};
\filldraw[black] (7,1.05) circle (1.5pt) node[anchor=west]{};
\filldraw[black] (5.48,0.5) circle (1.5pt) node[anchor=west]{};
\filldraw[black] (5,3.5) circle (1.5pt) node[anchor=west]{};
\filldraw[black] (10,0.5) circle (1.5pt) node[anchor=west]{};
\filldraw[black] (8.5,0.5) circle (1.5pt) node[anchor=west]{};

\end{tikzpicture}%
}
    \caption{\Gls{esl} design with the AEM40940 harvester and references to the sections where each component is elaborated.}%
    \label{fig:aem-esl}
\end{figure}

\textbf{Literature. }
\Glspl{esl} can be powered via an internal battery or externally via ambient or RF-based techniques. For example,~\cite{6615337,8889702} present battery-powered \gls{esl} designs with wireless communication system, achieving an autonomy of approximately \SI{3}{year}. Research focusing on energy harvesters using solar power and \SI{1}{\farad} \gls{edlc} buffer can be found in \cite{de2010design}.  However, this work focuses on RF-based \gls{wpt}.  Various RF \gls{wpt} techniques have been proposed to supply energy to low-power batteryless devices, as demonstrated in~\cite{10563994,10694523}. 
A recent study in \cite{wen2024915} on RF-powered \gls{esl} has primarily focused on voltage conversion techniques and efficiency optimization, with less emphasis on storage and peripheral device power consumption. 
Several commercial solutions claim advancements in RF-powered \gls{esl} technology~\cite{powercast_whitepaper,onio_shelf_label}. %
The white paper in~\cite{powercast_whitepaper}, introduces batteryless \gls{epd} solutions that requires a user to move through the store with a reader to update the \glspl{esl}, or alternatively, a robotic system could be deployed to automate this process. The authors claim a coverage range of \SI{24}{\meter}. However, the feasibility of this claim requires further evaluation, as at that distance the path loss exceeds \SI{60}{\dB}. \Gls{wpt} transmission at \SI{30}{\dBm} (as advertised) yields then a received power of only \SI{-30}{\dBm}, which falls below the sensitivity of the used harvester (\SI{-17}{\dBm})~\cite{pcc110_datasheet} and other conventional energy harvesters (\SI{-20}{\dBm})~\cite{7433469}. Consequently, \textit{achieving reliable energy transfer under these conditions remains highly challenging. This highlights the need for validated designs that are accessible and reproducible.}

\textbf{Contributions. } Our main contributions include the development of a novel RF-powered \gls{esl}, emphasizing the integration of \gls{cots} components to achieve a minimalistic and resource-efficient architecture. The design prioritizes the use of a low-cost and compact buffer capacitor. Furthermore, we experimentally validated and evaluated the system. This includes a detailed analysis of the \gls{esl}'s response time based on the energy consumption of the \gls{epd}, \gls{cots} harvester, and storage buffer. This investigation provides new insights into system performance and architectural design of RF-powered \glspl{esl}, as was lacking in prior research.

\textbf{Outline. }
The remainder of the paper is as follows. In \cref{sec:esl-arch}, we define the system requirements and present the proposed \gls{esl} architecture, detailing the selected microcontroller, harvester, and energy storage technologies based on the identified requirements and design constraints. \Cref{sec:exp} focuses on the experimental validation and evaluation of the \gls{esl}, including analyses of buffer charge time, efficiency, optimal \gls{rf} frequency, and optimal buffer capacity. In \cref{sec:tx}, we provide an overview of techniques to enhance the efficiency of \gls{wpt}. Finally, the paper concludes with a summary of findings and a discussion of potential future work in \cref{sec:concl}.

\section{ESL Requirements and Architecture}\label{sec:esl-arch}
This section outlines the system requirements for an RF-powered \gls{esl} capable of performing screen updates. Based on these requirements, the study evaluates an \gls{esl} built with \gls{cots} components, selecting a suitable \gls{epd} and microcontroller to control the screen. Additionally, the energy needed for an \gls{epd} update is measured, and the component design is analyzed considering constraints like buffer size, voltage ranges of the harvester, and booster converter. The final design is illustrated in~\cref{fig:aem-esl}.

\subsection{System Requirements and Design Constraints}

The system is designed to implement a functional Class~2 \gls{end} that operates efficiently under stringent constraints. The requirements (R) and assumptions (A) guiding the design are as follows:
\begin{itemize}
    \item[\req{1}] Minimize the energy buffer capacity to a single operation per full charge, i.e., an \gls{epd} update.
    \item[\req{2}] Power the device exclusively with an external \gls{rf} energy source.
    \item[\req{3}] Utilize \gls{cots} components to ensure cost-effectiveness and accessibility.
    \item[\req{4}] Optimize, i.e., minimize, the charge time to maximize system responsiveness.
    \item[\req{5}] Maximize receiver sensitivity, to charge at low \gls{rf} input powers.
    \item[\req{6}] Minimize the energy consumption of peripheral hardware.
    \item[\req{7}] Minimize the physical size of the \gls{esl}.
    \item[\textit{A1}] Impact of communication is not considered.
\end{itemize}

\subsection{RF Frequency Band Selection}
To address practical restrictions, a brief analysis is conducted to identify the most suitable frequency bands for RF-based \gls{wpt} of \glspl{esl} (\req{1}). Sub-1 GHz frequencies are preferred due to their lower attenuation compared to higher frequencies and the regulatory allowance for higher transmission powers in these bands. For instance, in Europe, a maximum transmit power of \SI{4}{\watt} (\SIrange[range-phrase={--},range-units=single]{915}{921}{MHz}) and \SI{2}{\watt} (\SIrange[range-phrase={--},range-units=single]{865}{868}{MHz}) is allowed~\cite{ETSIEN302208}.\footnote{Consult the regulations for more details on the specific bands and channel usage.}

\subsection{System Hardware Design}
Although it would be possible to directly charge the \gls{mcu} and \gls{epd} from the harvester, a storage element of buffer is added. This improves the sensitivity of the harvester, i.e., it lowers the minimum required \gls{rf} power on which the device can operate (\req{5}). To optimally use the energy buffer, the system needs to cope with a wide voltage range, when the buffer is being discharged. By doing so, the required capacity of the buffer can be reduced, minimizing both the initial charge time (\req{4}) and the required capacitance (\req{1}). It hence becomes essential to expand the dynamic range of the input voltage of the buffer and provide a stable output voltage to the \gls{mcu} and \gls{epd}.

\subsection{ePaper display and Microcontroller}\label{sec:mcu-epd}

In this study, an Adafruit evaluation board featuring a 2.13-inch monochrome eInk/ePaper display\footnote{Adafruit 2.13" eInk Display Breakouts and FeatherWings.} was selected. To refresh and communicate with the \gls{epd}, an ultra-low-power Arm Cortex-M0+ microcontroller is selected. The selected board operates at \SI{3.3}{V}, eliminating the need for level shifters between the display and \gls{mcu}. One update takes \SI{5.2}{\second} and requires \SI{75}{\milli\joule} as measured with the Otii Arc.

\subsection{Energy Storage Element and Boost Converter}\label{sec:cap-storage}
The capacitance of a capacitor is determined by the two voltage thresholds that define the operating range for updating the \gls{esl}. The maximum charging voltage, %
determined by the harvester, is referred to as the cut-off voltage $V_{\text{cutt-off}}$. Additionally, the minimum input voltage of the boost converter $V_{\text{boost, min}}$ determines the lowest usable voltage to which the capacitor can be discharged while still achieving the desired output voltage of \SI{3.3}{\volt}. Consequently, the capacitance can be calculated using:

\begin{equation}
    C_\text{B} = \frac{2 E_{\text{update}}}{V_{\text{cutt-off}}^2 - V_{\text{boost, min}}^2} \,.
    \label{eq:buffercapacitor}
\end{equation}

\subsection{RF Harvester -- Charging and Triggering Update}\label{sec:harvester}

The selected COTS harvester achieves efficiency levels of up to 30 percent, including tuning and \gls{pmu} efficiency~\cite{epeas_rf_aem40940}.
The \gls{pmu}, inside the harvester, (\req{3}) serves two primary functions in our design: charging the storage element (capacitor) and enabling the boost converter to wake up the microcontroller (\cref{fig:aem-esl}). 

The harvester is designed to accommodate various battery and capacitor technologies, with its configuration determining the voltage range for charging. This range is constrained by the overvoltage protection threshold, \( V_\text{ovch} \). The harvester includes two \glspl{ldo}, a low-voltage \\gls{ldo} and a high-voltage \gls{ldo}, to power peripherals using the attached storage element. Seven configurable settings define the overvoltage protection voltage, the activation and deactivation thresholds for the \glspl{ldo}, and the output voltages provided by both \glspl{ldo}. The choice of configuration depends on the storage element's technology.

To wake up the microcontroller when the capacitor is sufficiently charged (\( V_\text{chrdy} \)), the low-voltage \gls{ldo} activates the boost converter, which, in turn, powers the microcontroller. By utilizing the low-voltage \gls{ldo} to enable the boost converter, the need for additional hardware and energy consumption is significantly reduced (\req{6}). The ORing circuit, connected to one of the microcontroller's \glspl{gpio}, overrides the low-voltage \gls{ldo} output (\texttt{LVOUT}) when the capacitor voltage falls below \( V_\text{ovdis} \). Without this mechanism, the capacitor's usable voltage range would be limited to between \SI{2.8}{\volt} (\( V_\text{ovdis} \)) and \SI{3.1}{\volt} (\( V_\text{chrdy} \)), utilizing only \SI{18.4}{\percent} of its energy. By leveraging the full voltage range, the required capacitor capacity is significantly reduced (\req{1}, \req{4}).

Using \cref{eq:buffercapacitor} and taking into account the minimum voltage of the boost converter (\SI{0.9}{\volt}),
a minimum capacitor size of \SI{17}{\milli\farad} is calculated, allowing \SI{91.6}{\percent} of the capacitor's energy to be utilized. To account for additional losses and to follow the E12-series, a \SI{22}{\milli\farad} capacitor is used in this design.

\subsection{Capacitor Technology}\label{sec:cap-select}

Film, ceramic, tantalum, and polymer capacitors typically do not provide capacitance values in the millifarad range, are severely limited, or are prohibitively expensive. This in contrast to \glspl{edlc} and \glspl{aec}. While \glspl{aec} are available with capacitance values up to several tens of millifarads, their low energy density can lead to undesired large dimensions~(\req{7}). \Glspl{edlc}, with their higher energy density, present a promising alternative. \Cref{tab:storage-proposals} summarizes three options for this \gls{esl} application.

\begin{table}[ht]
    \centering
    \caption{Energy storage proposals considered during the selection process. (lower is better)}\label{tab:storage-proposals}
\begin{tblr}{
  colspec={l ccc},
  row{1}={font=\bfseries}, %
  width=0.8\linewidth, %
  cell{1-9}{4}={bg=ForestGreen!20}, %
  hline{1,4,10}={1pt}, %
  hline{9}={dashed}, %
  cells={font=\scriptsize} %
}
        Technology  & (1) EDLC & (2) EDLC   & (3) AEC \\
        Type  &  FYD0H223ZF & FYH0H105ZF & ECA-0JM223 \\
        Capacitance &  \SI{22}{\milli\farad} & \SI{1}{\farad} &\SI{22}{\milli\farad}\\
        Volume   & \SI{1.2}{cm^3}  &  \SI{7.0}{cm^3} &  \SI{9.0}{cm^3} \\
        Boost required &   Yes & No & Yes \\
        ESR &  \SI{200}{\ohm} & \SI{20}{\ohm} & \SI{0.05}{\ohm} \\
        Charge time (\cref{fig:charge_time_tuning})  & same & longer & same \\
        Cost &  $\approx$ \EUR\,3 & $\approx$ \EUR\,6 & $\approx$ \EUR\,1.5\\
        Usable &  No (ESR) &  Yes & Yes \\
\end{tblr}
\end{table}

The second option is similar to the first, with the significant difference that its capacitance is 45 times higher. This is required due to the high \gls{esr}, yielding high internal losses. 
This issue is no longer a concern, when increasing the capacity (Option 2). Additionally, a boost converter would not be required, as \SI{885}{\milli\joule} is available within the voltage range of \SIrange{2.8}{3.1}{\volt}. However, the initial charging time would become undesirably long, particularly when the buffer capacitor is fully discharged. Despite the slightly larger dimensions, the third option, proves to be the most suitable storage technology. Its low \gls{esr} ensures that internal losses are negligible. Additionally, its cost is generally lower than that of \glspl{edlc}. With the proposed architecture from \cref{fig:aem-esl}, the \gls{esl} can perform an update within \SIrange[range-units=single]{4}{120}{\minute} depending on the input power (\cref{fig:charge_time_tuning}).

\section{Experimental Evaluation and Validation}\label{sec:exp}

\begin{figure}[ht]
    \centering
    \includegraphics[width=0.8\linewidth]{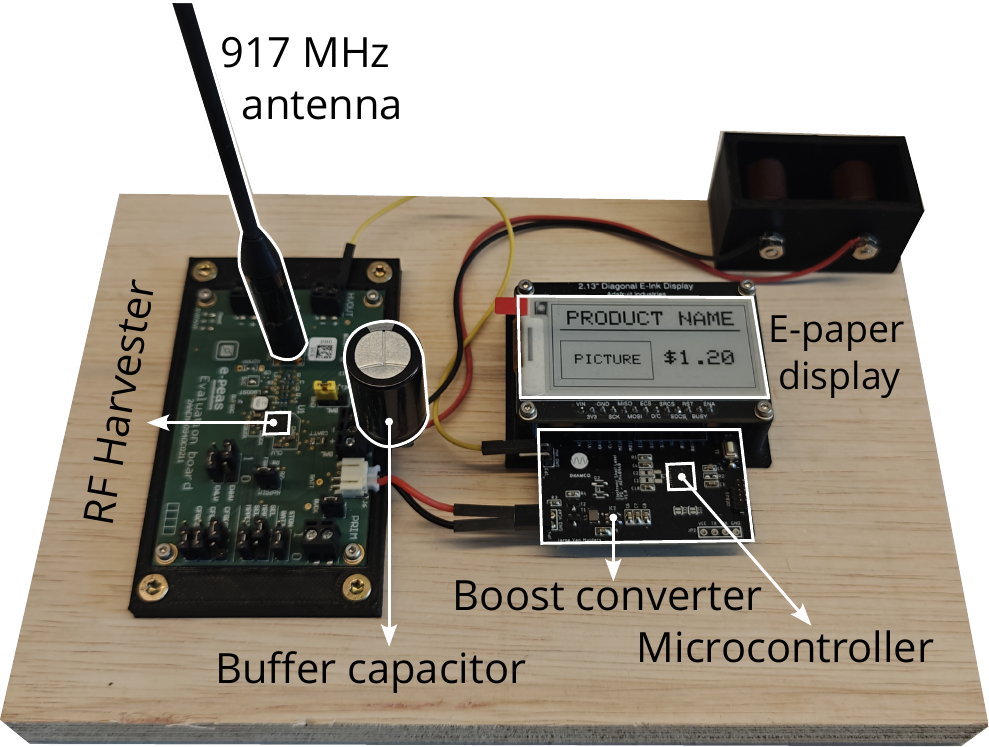}
    \caption{Prototype demonstration of the proposed \gls{esl} architecture. The black box in the upper right corner includes two connectors, allowing real-time voltage monitoring of the capacitor.}
    \label{fig:prototype}
\end{figure}

The implemented \gls{esl}, as presented in \cref{fig:prototype}, is evaluated by charging the selected \SI{22}{\milli\farad} capacitor (\cref{sec:cap-select}) using an RF generator (SMC100A) at \num{868} and \SI{920}{\mega\hertz}. The evaluation includes measuring the charge time (\cref{sec:exp:time}) and charge efficiency (\cref{sec:exp:efficiency}) under different input powers and \gls{rf} bands.

\subsection{Charge Time of the Buffer Capacitor}\label{sec:exp:time}

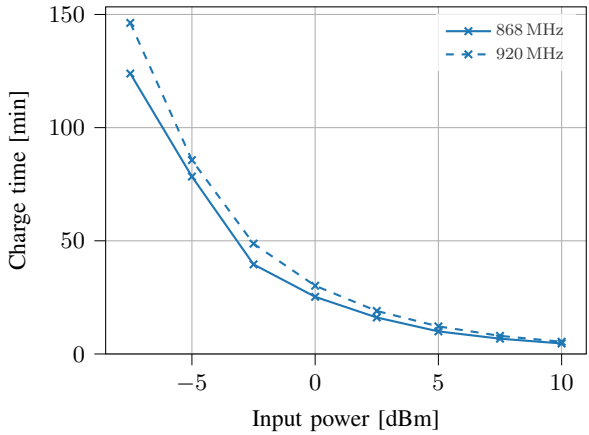
\begin{figure}[h]
    \centering
    \begin{tikzpicture}

\definecolor{darkgray176}{RGB}{176,176,176}
\definecolor{darkorange25512714}{RGB}{255,127,14}
\definecolor{lightgray204}{RGB}{204,204,204}
\definecolor{steelblue31119180}{RGB}{31,119,180}

\begin{axis}[
width = 0.9\linewidth,
height = 0.7\linewidth,
legend cell align={left},
legend style={fill opacity=0.8},
tick align=outside,
tick pos=left,
x grid style={darkgray176},
xlabel={Input power [dBm]},
xmajorgrids,
xmin=-8.5, xmax=11,
xtick style={color=black},
y grid style={darkgray176},
ylabel={Charge time [min]},
ymajorgrids,
ymin=0, ymax=153.363366731803,
ytick style={color=black}
]
\addplot [
    color= steelblue31119180,
    mark=x,
    thick,
    solid,
    mark options={solid}, mark size=2pt
]
table {%
-7.5 123.901360988617
-5 78.3801770369212
-2.5 39.5771076321602
0 25.2549402832985
2.5 16.1140583753586
5 9.97545903921127
7.5 6.75046074787776
10 4.64821739594142
};
\addlegendentry{\SI{868}{MHz}}
\addplot [
    color=steelblue31119180,
    mark=x,
    dashed,
    thick,
    mark options={solid}, mark size=2pt,
]
table {%
-7.5 146.281692953904
-5 85.6366419037183
-2.5 48.7331524133682
0 30.1430814385414
2.5 18.9945779403051
5 12.1891778945923
7.5 7.99466664791107
10 5.38257083495458
};
\addlegendentry{\SI{920}{MHz}}
\end{axis}

\end{tikzpicture}
    \caption{Measured time to charge the buffer capacitor of \SI{22}{\milli\farad} to \SI{3.1}{\volt} with the AEM40940 harvester.}
    \label{fig:charge_time_tuning}
\end{figure}

\Cref{fig:charge_time_tuning} illustrates the measured charging time of a \SI{22}{\milli\farad} \gls{aec} up to a voltage of \SI{3.1}{\volt}. 
The harvester exhibits slightly improved performance at a frequency of \SI{868}{\mega\hertz}. Specifically, for lower input voltage levels, the charging time at \SI{868}{\mega\hertz} is reduced by several minutes compared to \SI{920}{\mega\hertz}, given the same input power level. The reason is that harvester itself is optimally tuned for \SI{868}{\mega\hertz}.

\subsection{Harvester Charge Efficiency}\label{sec:exp:efficiency}

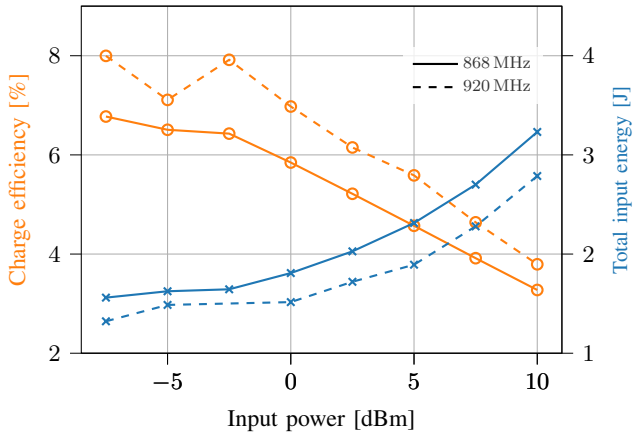
\begin{figure}[h]
    \centering
    \begin{tikzpicture}

\definecolor{darkgray176}{RGB}{176,176,176}
\definecolor{darkorange25512714}{RGB}{255,127,14}
\definecolor{lightgray204}{RGB}{204,204,204}
\definecolor{gray100}{RGB}{100,100,100}
\definecolor{steelblue31119180}{RGB}{31,119,180}

\begin{axis}[
    height=0.7\linewidth,
    width=0.9\linewidth,
legend cell align={left},
legend style={
  fill opacity=0.8,
  draw opacity=1,
  text opacity=1,
  at={(0.5,0.8)},
  anchor=north west,
  draw=lightgray204
},
tick align=outside,
tick pos=left,
x grid style={darkgray176},
xlabel={Input power [dBm]},
xmajorgrids=true,
xmin=-8.5, xmax=11,
xtick style={color=black},
y grid style={darkgray176},
ylabel={\color{darkorange25512714} Charge efficiency [\%]},
ymajorgrids=true,
ymin=2, ymax=9,
ytick style={color=black}
]
\addplot [
    color= darkorange25512714,
    mark=o,
    thick,
    dashed,
    mark options={solid}, mark size=2pt, forget plot
]
table {%
-7.5 7.99720672178668
-5 7.10934515569659
-2.5 7.91747535247243
0 6.97577595885733
2.5 6.14986765872612
5 5.58687746910244
7.5 4.63988597729201
10 3.79358694019815
};
\addplot [
    color=darkorange25512714,
    mark=o,
    thick,
    mark options={solid}, mark size=2pt, forget plot
]
table {%
-7.5 6.7736157221062
-5 6.50644541029938
-2.5 6.42908098090982
0 5.84581821909314
2.5 5.21567305219707
5 4.57017031128929
7.5 3.91689865027037
10 3.27552593364163
};
\end{axis}

\begin{axis}[
    height=0.7\linewidth,
    width=0.9\linewidth,
axis y line=right,
legend style={
  at={(0.67,0.80)},
  anchor=west,
},
tick align=outside,
x grid style={darkgray176},
xmin=-8.5, xmax=11,
xtick pos=left,
xtick style={color=black},
y grid style={darkgray176},
ylabel={\color{steelblue31119180} Total input energy [J]},
ymin=1, ymax=4.5,
ytick pos=right,
ytick style={color=black},
yticklabel style={anchor=west},
xmajorgrids=false,
ymajorgrids=false,
axis line style={-}
]
\addplot [
    color= steelblue31119180,
    mark=x,
    thick,
    dashed,
    mark options={solid}, mark size=2pt, forget plot
]
table {%
-7.5 1.32194302870767
-5 1.48715590641858
0 1.51550462627411
2.5 1.72028056759094
5 1.89271810439604
7.5 2.2793862049246
10 2.78698872089386
};

\addplot [
    color= steelblue31119180,
    mark=x,
    thick,
    mark options={solid}, mark size=2pt, forget plot
]
table {%
-7.5 1.56073358486158
-5 1.62482496209878
-2.5 1.64463859101075
0 1.80844052171707
2.5 2.02709634747836
5 2.31322311969105
7.5 2.70011823661367
10 3.23116632461548
};

\addlegendimage{black, solid,thick}
\addlegendimage{black, dashed, thick}
\addlegendentry{\SI{868}{MHz}}
\addlegendentry{\SI{920}{MHz}}
\end{axis}

\end{tikzpicture}
    \caption{Measured efficiency related to the RF input energy and the stored energy in the buffer.}
    \label{fig:charge_efficiency}
\end{figure}

As shown in \cref{fig:charge_time_tuning}, charging logically occurs much faster at higher input power levels. However, this is not necessarily the most energy-efficient. The required input \gls{rf} \emph{energy} at the harvester is significantly higher at higher \gls{rf} input \emph{power} levels, as illustrated in \cref{fig:charge_efficiency}. Hence, the charge efficiency, i.e., the ratio of required input energy to the nominal energy of the buffer, decreases with input \gls{rf} power. The efficiency naturally takes into account the tuning losses, the harvester AC/DC conversion and the \gls{pmu} losses to charge the capacitor.

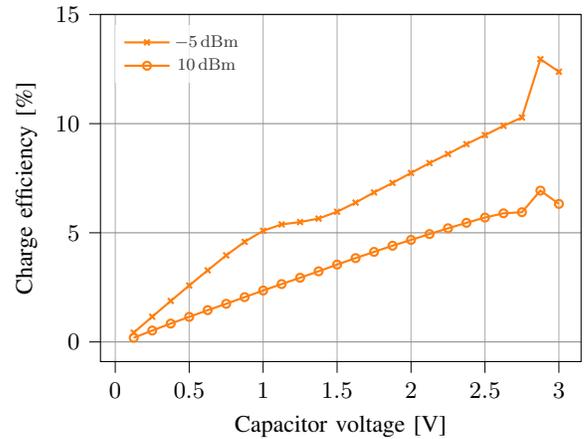
\begin{figure}[h]
    \centering
    \begin{tikzpicture}

\definecolor{darkgray176}{RGB}{176,176,176}
\definecolor{darkorange25512714}{RGB}{255,127,14}

\begin{axis}[
    height=0.7\linewidth,
    width=0.9\linewidth,
    tick align=outside,
    tick pos=left,
    x grid style={darkgray176},
    xlabel={Capacitor voltage [V]},
    ylabel={Charge efficiency [\%]},
    xmin=-0.1, xmax=3.15,
    ymin=-0.9, ymax=15,
    xtick style={color=black},
    ytick style={color=black},
    legend pos=north west,
]
\addplot[
    color=darkorange25512714,
    mark=x,
    thick,
    mark options={solid}, mark size=1.5pt,
] coordinates {
(0.125,0.428298634962978)
(0.25,1.15440254145031)
(0.375,1.87665213771732)
(0.5,2.58404915625955)
(0.625,3.28025015092578)
(0.75,3.96436953632097)
(0.875,4.5860301309526)
(1,5.09162445857025)
(1.125,5.38308278858274)
(1.25,5.49122732060493)
(1.375,5.65255396793768)
(1.5,5.96265152988152)
(1.625,6.3817399866605)
(1.75,6.848350562132)
(1.875,7.28438604657872)
(2,7.74144972764007)
(2.125,8.19447780825329)
(2.25,8.61090602121549)
(2.375,9.06149384280822)
(2.5,9.47584164178325)
(2.625,9.89677807379963)
(2.75,10.2791421714723)
(2.875,12.9523728345608)
(3,12.3772424211308)
};

\addplot[
    color= darkorange25512714,
    mark=o,
    thick,
    mark options={solid}, mark size=1.5pt
] coordinates {
(0.125,0.190539308106177)
(0.25,0.5224581833378)
(0.375,0.840459998601059)
(0.5,1.1482255420802)
(0.625,1.45566965071456)
(0.75,1.74581675388377)
(0.875,2.05158004481653)
(1,2.35445880106922)
(1.125,2.65015381080957)
(1.25,2.93998251900657)
(1.375,3.23486374921233)
(1.5,3.54318558568533)
(1.625,3.83884755783891)
(1.75,4.12467720635155)
(1.875,4.40773631237346)
(2,4.67336702638427)
(2.125,4.94093814722085)
(2.25,5.20552331493076)
(2.375,5.45584743400769)
(2.5,5.6970791701015)
(2.625,5.89212730607576)
(2.75,5.94156609213158)
(2.875,6.93200193910745)
(3,6.32324320409129)
};

\legend{\SI{-5}{\dBm}, \SI{10}{\dBm}}
\end{axis}

\end{tikzpicture}
    \caption{Measured harvester efficiency at \SI{868}{\mega\hertz} as a function of the capacitor voltage for input power levels of \SI{-5}{\dBm} and \SI{10}{\dBm}.}%
    \label{fig:charge_eff_cap_voltage}
\end{figure}

The charging efficiency depends, among others, on the initial voltage over the capacitor. In~\cref{fig:charge_eff_cap_voltage}, the measured charge efficiency is depicted with respect to this initial capacitor voltage. As can be observed, the higher the initial voltage, the higher the charge efficiency. Only considering this, a higher capacity would be desirable as it would be operating at a higher voltage and would consequently require a lower voltage range. For example,  a \SI{1}{\farad} (option~2 in \cref{tab:storage-proposals}) \gls{edlc} capacitor, would have sufficient energy in the voltage range \SIrange{3.0}{3.1}{\volt}. Furthermore, this would mean that no boost converter would be necessary. The downside of the higher capacity is the increased initial charge time, a higher \gls{esr} ($\times 400$), increased size (depending on the technology) and cost ($\times 4$).

\section{Techniques to Deliver Energy to ESLs}\label{sec:tx}
The transmit side is also important to consider, although this is not the focus of this paper, and it does not impact the findings and conclusions made here. Various RF-based strategies and technologies can be utilized to deliver sufficient power to the \gls{esl}. Power transfer can be achieved using either a single antenna or multiple antennas, with the latter offering, among others, higher received power and improved transfer efficiency compared to a single-antenna setup~\cite{10694523}.

Recent studies have investigated leveraging environmental information to coherently combine RF signals at the \glspl{esl}, relying solely on geometric details of the environment, \gls{esl}, and antenna positions~\cite{DeutschmannICC2022,Call2503:Experimental}. Beyond geometry-based methods, reciprocity-based approaches use an uplink pilot to learn the downlink channel from the uplink channel, as discussed in~\cite{Call2503:Experimental}. However, both approaches have limitations: geometry-based methods require accurate positional information, and reciprocity-based techniques depend on sufficient initial on-board energy to transmit the pilot signal. 
To address these initial access challenges, several alternative techniques have been proposed in the literature, such as random-phase sweeping and high-\gls{papr} transmission~\cite{DeutschmannICC2022,clerckx2018beneficial,142775,VanM2410:Single}. 

\textit{The multi-antenna coherent \gls{wpt} system is demonstrated using our testbed, receiving \SI{0}{\dBm} at the \gls{esl}.~\cite{callebaut2024experimentalstudyeffectsynchronization} This enables the screen to update every \SI{25}{\minute}.} %

\balance%
\section{Conclusion and Future Work}\label{sec:concl}
This preliminary study demonstrates the feasibility of a batteryless \gls{esl} system powered by RF-\gls{wpt}. It is validated through our testbed and detailed analysis of design choices and architectural decisions are presented here. Despite challenges such as the \gls{esl} energy consumption and low harvester efficiency%
, this work highlights the potential for significant improvements in future iterations. The system was optimized to leverage a leaded aluminum electrolytic capacitor, balancing cost and performance, though its limited capacitance and physical size suggest that cascading buffers or transitioning to EDLCs may be necessary for higher power applications. Future work should focus on enhancing harvester sensitivity, reducing capacitor sizes, improving antenna design, and lowering operating voltages to align with component specifications. Additionally, multi-antenna \gls{wpt} and initial access schemes will be investigated to further improve the overall efficiency and feasibility of next-generation \glspl{esl}.
The findings and results of this work lay the groundwork for further optimizations in hardware resource efficiency and energy consumption, advancing the practicality of RF-powered ESL systems.

\printbibliography

\end{document}